\documentclass[11pt,twoside]{article}
\usepackage[pdftex]{graphicx}

\usepackage{amsmath}
\usepackage{amssymb}
\usepackage{cite}

\begin{document}
\newcommand{\pst}{\hspace*{1.5em}}

\newcommand{\rigmark}{\em Journal of Russian Laser Research}
\newcommand{\lemark}{\em Volume 30, Number 5, 2009}

\newcommand{\be}{\begin{equation}}
\newcommand{\ee}{\end{equation}}
\newcommand{\bm}{\boldmath}
\newcommand{\ds}{\displaystyle}
\newcommand{\bea}{\begin{eqnarray}}
\newcommand{\eea}{\end{eqnarray}}
\newcommand{\ba}{\begin{array}}
\newcommand{\ea}{\end{array}}
\newcommand{\arcsinh}{\mathop{\rm arcsinh}\nolimits}
\newcommand{\arctanh}{\mathop{\rm arctanh}\nolimits}
\newcommand{\bc}{\begin{center}}
\newcommand{\ec}{\end{center}}

\thispagestyle{plain}

\label{sh}


\begin{center} {\Large \bf
\begin{tabular}{c}
Bound state of particle in delta-potential
\\[-1mm]
in tomographic representation of quantum mechanics
\end{tabular}
 } \end{center}

\bigskip

\bigskip

\begin{center} {\bf
 I.V.Dudinetc$^{1*}$ and V.I.Manko$^2$
}\end{center}

\medskip

\begin{center}
{\it
$^1$Moscow Institute of Physics and Technology  \\
Institutskii per.9,Dolgoprudnyi,Moscow Region 141700,Russia\\

\smallskip

$^2$Lebedev Physical Institute,Russian Academy of Sciences \\
Leninskii Prospect 53,Moscow 119991,Russia}
\smallskip
\end{center}
\begin{abstract}\noindent
The problem of quantum particle moving in Dirac delta potential with instant changing well depth is
studied by using formalism of tomographic representation of quantum mechanics.The bound state tomogram is given in terms of error
function.The ionisation probability due to instant change of the potential parameter is calculated in terms of integral containing the state tomograms
.The probability of ionisation is also expressed in terms of the Wigner function.
\end{abstract}

\medskip


\section{Introduction}
\pst

The behaviour of quantum particle moving in delta-potential was considered e.g. in \cite{1} and the wave function was obtained in explicite form.
Recently the tomographic picture of quantum
mechanics was introduced in \cite{2,3}.The aim of this work is to obtain the results of studying the quantum motion of particle in delta potential in
formalism of tomographic representation of quantum mechanics.Some aspects of this problem were considered in \cite{4}.In this work we will repeat and correct some calculations.
The tomographic probability distribution called symplectic tomogram of a quantum state is connected with the wave function of the pure state by
integral transform related to Fresnel integral \cite{5,6}.The mixed state has the tomogram which can be obtained from the Wigner function $W(q,p)$
introduced in \cite{7} by means of the Radon integral transform \cite{8}.There exist different tomographic probability distributions which can be identified
with the quantum states (see e.g.\cite{3}).The optical tomogram of the quantum state was introduced in \cite{9,10}.The symplectic tomogram
of the quantum state was introduced in \cite{11}.The suggestion to identify the quantum states with the tomographic probability distributions instead of
wave functions \cite{12} or density matrices\cite{13,14}was done in explicit form in \cite{2}.
Since the tomographic probability distributions are connected with wave functions or density matrices by invertible integral transform the quantum mechanical
equations like Shroedinger equations for energy levels or for the wave function evolution can be rewritten as the equations for the state tomograms.
Such equations were obtained in \cite{2} for symplectic tomogram and for optical tomogram the equations were given in \cite{15,16}.
Thus,the tomographic picture of quantum mechanics from the mathematical point of view is another representation of quantum mechanical equations.The
main advantage of the equations is that they are written for fair probability distributions replacing complex wave functions or density matrices.It is clear that
all solved problems of quantum mechanics can be reconsidered in the new probability represantation .This provides a possibility to study some new aspects of
quantum states though all these aspects are contained in usual wave function or density matrix representations.Nevertheless the information properties
like Shannon entropy \cite{17} or Renyi entropy \cite{18} associated with the quantum states can be elucidated easier in the tomographic probability representation
\cite{19}.On the other hand the solving the analogs of Shroedinger equation for energy levels for some potentials provides in the tomographic represantation extra difficulties.
We will study for the concrete potential energy the connection of Shroedinger equation and its corresponding solutions to the equations for tomograms.
One of our goals is to consider the relations(for concrete potential energy) of solutions of the  standart Shroedinger equation with its tomographic analog.
\\We will consider the Dirac delta-potential in the Hamiltonian of our system as the particular example.
The paper is organized as follows.
In Sec. 2 we calculate symplectic and optical tomograms for bound state of particle moving in delta potential.
In Sec. 3 we obtain the Wigner function for the particle in  bound state of delta potential.
In Sec. 4 we consider two delta potentials.We calculate the Wigner function and get formula for tomogram.
In Sec. 5 we review problem of shaking.
In Sec.6 we calculate momenta $<x^2 >$ and $<p^2 > $ using the tomographic approach to description of quantum states.
In Sec.7 the conclusions and prospectives are presented.
\section{Symplectic tomogram for one delta-potential}
\pst
Let us consider a particle moving in  delta-potential of the form
\begin{equation}
V(x)=-\chi\delta (x).
\end{equation}
The real parameter $\chi$ describes the interaction force of the particle with the potential.
It is known \cite{1} that the wave function of the bound state of the particle in  delta-potential reads
\begin{equation}
\psi (y)=\sqrt{\chi}e^{-\chi|y|}.
\end{equation}
Below we take the Plank constant  $\hbar$ and mass $m$ of the particle to be equal to unity($\hbar$=$m$=1).
The symplectic tomogram $M(X,\mu ,\nu )$ can be found from formula(see \cite{20})
\begin{equation}
M(X,\mu ,\nu )=\frac{1 }{2\pi |\nu|}\left | \int_{-\infty }^{\infty }\psi (y)e^{\frac{i\mu}{2\nu }y^2-\frac{iX}{\nu}y}dy \right | ^{2}.
\end{equation}
Here $\mu$ and $\nu$ are real parameters,which describe an ensemble of rotated and scaled reference frames,X is the particle position
measured in the rotated and scaled reference frame in the particle phase-space.
\\Symplectic tomogram for bound state of particle in delta-potential can be given by direct calculation in view of (3)

\begin{equation}
M(X,\mu ,\nu )=\frac{\chi }{4 |\mu|} \left |e^{-\frac{\chi X}{\mu}}erfc \left \{ \frac{-X+i\chi \nu }{\sqrt{2\mu\nu i}}\right \}+
e^{\frac{\chi X}{\mu}}erfc \left \{ \frac{X+i\chi \nu }{\sqrt{2\mu\nu i}}\right \}  \right | ^2,
\end{equation}
where we used the known integral
\begin{equation}
 \int_{0}^{\infty }e^{-z^2+2kz}dz=\frac{\sqrt{\pi }}{2}e^{k^2}erfc(k)
\end{equation}
and error function is defined as
\begin{equation}
 erfc(x)=\frac{2}{\sqrt{\pi}}\int_{x}^{\infty }e^{-t^2}dt.
\end{equation}
\\There is normalization condition for the tomogram(see \cite{19}),which is valid for normalized wave function
\begin{equation}
\int_{-\infty }^{\infty } M(X,\mu,\nu)dX=1.
\end{equation}
Using the formula for symplectic tomogram (4) and the normalization condition (7),we have
\begin{equation}
\frac{\chi}{2|\mu|}\int_{-\infty }^{\infty }e^{-\frac{2\chi}{\mu}X}\left | erfc\left \{ \frac{-X+i\chi \nu }{\sqrt{2\mu\nu i}} \right \} \right |^2dX=1.
\end{equation}
We used the known integral
\begin{equation}
\int_{0}^{\infty }erfc(ax)erfc(bx)dx=\frac{a+b-\sqrt{a^2+b^2}}{ab\sqrt{\pi}}.
\end{equation}
For $\mu =\cos \theta $ and $\nu =\sin \theta $,the symplectic tomogram $M(X,\mu ,\nu )$ is the probability distribution for the homodyne variable
used in optical tomography \cite{10,21}.Substituting the expressions $\mu =\cos \theta $ and $\nu =\sin \theta $ into (4) we have optical tomogram in explicite form

\begin{equation}
w(X,\theta)=M(X,\cos \theta ,\sin \theta )=\frac{\chi }{4 |\cos \theta|} \left |e^{-\frac{\chi X}{\cos \theta}}erfc \left \{ \frac{-X+i \chi \sin \theta  }{\sqrt{i\sin 2\theta}}\right \}+
e^{\frac{\chi X}{\cos \theta}}erfc \left \{ \frac{X+i \chi \sin \theta }{\sqrt{i\sin 2\theta }}\right \}  \right | ^2.
\end{equation}

\section{The Wigner function for  delta-potential}
\pst
Our aim in this section  is to discuss some features of bound state of particle moving in delta-potential,discussed in previous section,
in terms of the Wigner function.

The Wigner function $W(q,p)$ determines completely the quantum state of a system(see e.g.\cite{21}),and it
is expressed in terms of the wave function as
\begin{equation}
 W(q,p)=\int du e^{-ipu}\psi ^{*}(q+\frac{u}{2})\psi (q-\frac{u}{2}).
\end{equation}
The configuration space wave function is denoted by $\psi(x)$ and $p$ is the momentum.
The symplectic tomogram $M(X,\mu,\nu)$ is related to the state of the quantum system expressed in terms of the Wigner function by means of the integral transform
\begin{equation}
M(X,\mu,\nu)=\int W(q,p)\delta (X-\mu q - \nu p)\frac{dqdp}{2\pi }.
\end{equation}
(see \cite{3,20})
The Wigner function is completely determined by the symplectic tomogram $M(X,\mu,\nu)$ and vice versa.
We say that quantum state is given if the position probability distribution $M(X,\mu,\nu)$ in an ensemble of rotated and scaled reference
frames in classical phase space is given(see \cite{21}).
\\ The Wigner function for particle moving in delta-potential can be found from (2) and (11),and

the result for the Wigner function reads(\cite{4})
\begin{equation}
W(q,p)=\frac{2\chi^2e^{-2|q|\chi}}{p^2+\chi^2}\left \{ \cos(2p|q|)+\frac{\chi}{p}\sin(2p|q|) \right \}.
\end{equation}

\begin{figure}[ht]
\bc \includegraphics[width=8.6cm]{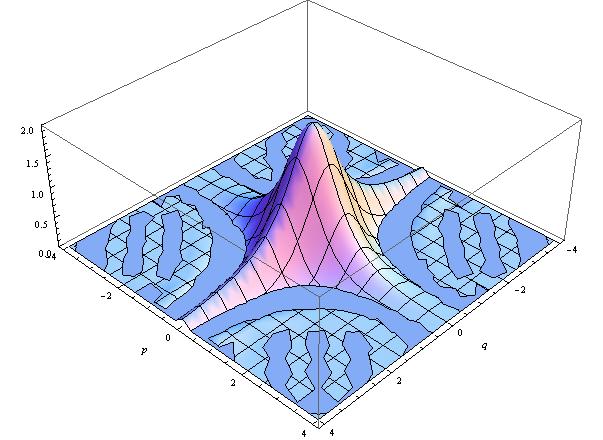}
  \ec
\vspace{-4mm}
\caption{
The Wigner function $W(q,p)$ (13) for particle,moving in delta potential $V(x)=-\chi\delta (x)$ for $\chi=1$. }
\end{figure}

The Wigner function has property
\begin{equation}
\frac{1}{2\pi}\int_{-\infty }^{\infty } W(x,p)dp=\left | \Psi(x) \right |^2,
\end{equation}
which follows from definition of the Wigner function (11) and the delta function Fourier representation
$$\frac{1}{2\pi}\int_{-\infty }^{\infty } e^{ikx} dx=\delta(k).$$
The Wigner function (13)  satisfies equation (14).
Using (14) we get
\begin{equation}
\int W(q,p)\frac{dqdp}{2\pi}=\int dq \left | \Psi(q) \right |^2=1.
\end{equation}
Symplectic tomogram $M(X,\mu,\nu)$ can be calculated from the Wigner function $W(q,p)$ (see formula (12)).
\begin{equation}
M(X,\mu,\nu)=\frac{1}{2\pi|\nu|}\int W(q,\frac{X-\mu q}{\nu})dq.
\end{equation}
The symplectic tomogram is expressed in terms of the integral
\begin{equation}
M(X,\mu,\nu)=\frac{\chi^2}{\pi|\nu|}\int dq \frac{e^{-2|q|\chi}}{p(q)^2+\chi^2}\left \{ \cos(2p(q)|q|)+\frac{\chi}{p(q)}\sin(2p(q)|q|) \right \},					 
\end{equation}
where $p(q)=\frac{X-\mu q}{\nu}$ and tomogram $M(X,\mu,\nu)$ is defined by the formula(4).
We managed numerically to verify (17).

\section{Two delta potentials}
\pst
In this section we calculate the Wigner function for bound states of particle moving in  two delta-potentials.
\\Let consider two delta potentials with one parameter $\chi$
\begin{equation}
V(x)=-\chi\delta (x-a)-\chi\delta (x+a).
\end{equation}
There are two(symmetric and asymmetric) solutions of Shrodinger equation for two $\delta$-potentials,which correspond to two bound states
\begin{equation}
{\psi _ \pm }(x) = \left\{ \begin{array}{l}
 \pm B{e^{\beta x}}\\
C({e^{\beta x}} \pm {e^{ - \beta x}})\\
B{e^{ - \beta x}}
\end{array} \right\}\\.
\end{equation}
Formula (19) can be represented in the form(see \cite{4})
\begin{equation}
{\psi _ \pm }(x) = C(e^{-\beta|x-a|}\pm e^{-\beta|x+a|}),
\end{equation}
where C corresponds to normalization condition for the wave function $\Psi(x)$
\begin{equation}
\int_{-\infty }^{\infty }|{\psi _ \pm }(x)|^2dx=1.
\end{equation}
For $\psi_{+}(x)$ parameter $\beta$ satisfies relation $\chi=\frac{\beta}{1+e^{-2\beta a}}$ and this symmetric solution always exists.
And for $\psi_{-}(x)$ we have $\chi=\frac{\beta}{1-e^{-2\beta a}}$.Antisymmetric solution exists if $\chi>\frac{1}{2a}$.

Let us define the function
\begin{equation}
F(p,q)\equiv \int_{-\infty }^{\infty }e^{-ipu}e^{-\beta|q+\frac{u}{2}|}e^{-\beta |q-\frac{u}{2}|}.
\end{equation}

Analogically to derivation of the Wigner function in previous section we get the formula
\begin{equation}
\\F(p,q)=2e^{-2\beta|q|}[\frac{\sin(2p|q|)}{p}+\frac{\beta \cos(2p|q|)}{p^2+\beta^2}-\frac{p\sin(2p|q|)}{p^2+\beta^2}].
\end{equation}
The Wigner function $W(q,p)$ can be expressed using the above function $F(p,q)$(\cite{21}).

The Wigner function is
\begin{equation}
W_{\pm}(q,p)=C^2(F(p,q-a)\pm 2\cos(2pa)F(p,q)+F(p,q+a) ).
\end{equation}
On figure 2 we illustrated symmetric $W_{+}(q,p)$ and asymmetric $W_{-}(q,p)$ the Wigner functions for particle moving in the potential $V(x)$ (18) for
 parameter $a=0.5$ and for $C=1$,$\chi=1$.For $a \to 0$ the Wigner function $W_{+}(q,p)$ agrees with (13) and $W_{-}(q,p) \to 0$.For
increasing values of $a$ ,shapes of the Wigner functions $W_{+}(q,p)$ and  $W_{-}(q,p)$ change from smooth to increasing peaks.The Wigner function is analogue of
distribution function.But unlike classical distribution function,the Wigner function can be negative.On figure 3 we can see,that the Wigner function has negative values.
Tomogram can be found  by using formula (12) and the homogeneity property  of delta-function $\delta(ax)=\frac{1}{|a|}\delta(x) $

The result is(see \cite{21})

$$M_{\pm}(X,\mu,\nu)=\frac{C^2}{2\pi |\nu|}\int dq F(\frac{x-\mu (q+a)}{\nu},q)
 \pm \frac{C^2}{2\pi |\nu|}\int dq F(\frac{x-\mu q}{\nu},q)2\cos(2a\frac{x-\mu q}{\nu})+$$
\begin{equation}
 +\frac{C^2}{2\pi |\nu|}\int dq F(\frac{x-\mu (q-a)}{\nu},q).
\end{equation}

\begin{figure}[ht]
\bc \includegraphics[width=8.6cm]{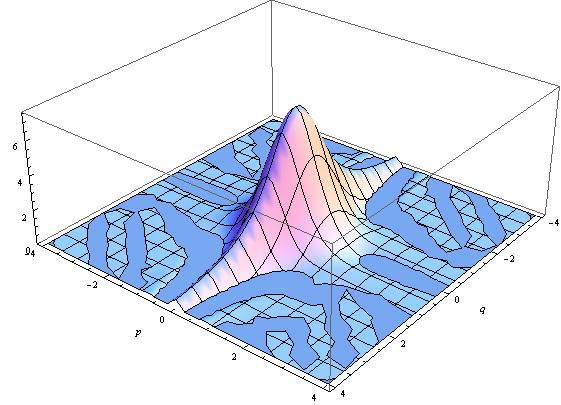}
 \includegraphics[width=8.6cm]{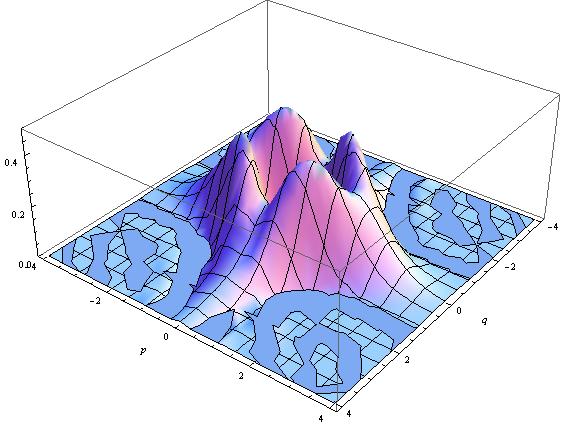}
  \ec
\vspace{-4mm}
\caption{
The Wigner functions $W_{+}(q,p)$ (fig. 2(a))and $W_{-}(q,p)$ (fig.2(b))  for particle,moving in two potentials $V(x)=-\chi\delta (x-a)-\chi\delta (x+a) $ for $C=1$, $\chi=1$ and $a=0.5$.}
\end{figure}

\begin{figure}[ht]
\bc \includegraphics[width=8.6cm]{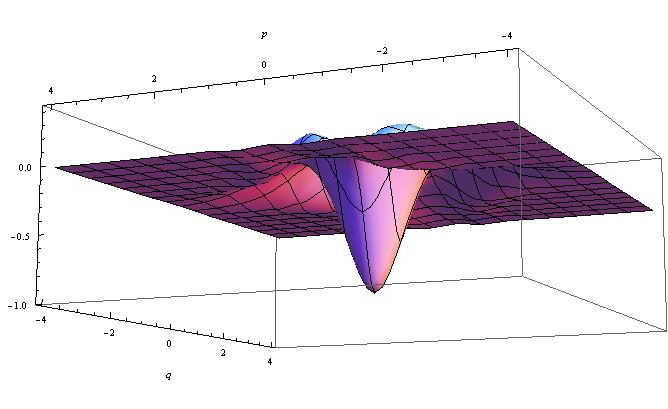}

  \ec
\vspace{-4mm}
\caption{
The Wigner functions $W_{-}(q,p)$  for particle,moving in two potentials $V(x)=-\chi\delta (x-a)-\chi\delta (x+a) $ for $C=1$, $\chi=1$ and $a=0.5$.}
\end{figure}

\section{Problem of shaking}
\pst
In this section we will review quickly changing of parameter of delta-potential.We will calculate probability to remain in bound state
through the Wigner function.
\\We have $\delta$-potential
$$ V_{1}(x)=-\kappa_{1} \delta(x)$$
with one parameter $\kappa_{1}$.
After that $\kappa_{1}$ changes instantly to $\kappa_{2}$
$$ V_{2}(x)=-\kappa_{2} \delta(x).$$
Wave functions of the bound state of the particle,moving in this delta potentials are
$\psi_{1} (x)=\sqrt{\kappa_{1}}e^{-\kappa_{1}|x|}$
$\psi_{2} (x)=\sqrt{\kappa_{2}}e^{-\kappa_{2}|x|}.$

Probability to remain in a bound state can be found using wave function
\begin{equation}
 P(0 \to 1)=|\int \psi^{*}_{1}(x) \psi_{2} (x)dx|^{2}.
\end{equation}
The result is
\begin{equation}
 P(0 \to 1)=\frac{4\kappa_{1}\kappa_{2}}{(\kappa_{1}+\kappa_{2})^{2}}.
\end{equation}

Quantum-transition probability can be expressed in terms of an overlap integral of  symplectic tomograms.
If the pure state of  a quantum system is described by symplectic tomogram
$M_{1}(X,\nu,\mu)$
and the final state of the quantum system is described by the marginal distribution
$M_{2}(X,\nu,\mu)$,
the probability of the quantum transition $P(0 \to 1)$ can be expressed in terms of their overlap integral.Also it can be given in terms  of overlap integral of the Wigner functions $W_{1}(q,p)$ and $W_{2}(q,p)$ (see \cite{20}).
The Wigner functions $W_{1}(q,p)$ and $W_{2}(q,p)$ of the bound state of the particle moving in delta potentials $V_{1}$ and $V_{2}$  can be found from wave functions and definition of the Wigner function (11)
(see Appendix 1)
\begin{equation}
P(0 \to 1)= \frac{1}{2\pi}\int W_{1}(q,p)W_{2}(q,p)dqdp.
\end{equation}

Using (13) and (28) we can get (27).
The transition probability $P(0 \to 1)$ can be expressed in view of the symplectic tomograms(see Appendix 2)
\begin{equation}
P(0 \to 1)= \int M_{1}(X,\mu ,\nu )M_{2}(Y,-\mu ,-\nu )e^{i(X+Y)}\frac{dXdY d \mu d\nu}{2\pi}.
\end{equation}

If $\kappa_{1}=\kappa_{2}$ then $P(0 \to 1)=1$ and we have (\cite{21})
\begin{equation}
\int M_{1}(X,\mu ,\nu )M_{1}(Y,-\mu ,-\nu )e^{i(X+Y)}\frac{dXdY d \mu d\nu}{2\pi}=1.
\end{equation}
Using(4) we have
$$\frac{\chi^2 }{16|\mu|}\int  \left |  e^{-\frac{\chi X}{\mu}}erfc \left \{ \frac{-X+i\chi \nu }{\sqrt{2\mu\nu i}}\right \}+e^{\frac{\chi X}{\mu}}erfc \left \{ \frac{X+i\chi \nu }{\sqrt{2\mu\nu i}}\right \}  \right | ^2$$
$$\left |  e^{\frac{\chi Y}{\mu}}erfc \left \{ \frac{-Y-i\chi \nu }{\sqrt{2\mu\nu i}}\right \}+e^{-\frac{\chi Y}{\mu}}erfc \left \{ \frac{Y-i\chi \nu }{\sqrt{2\mu\nu i}}\right \}  \right | ^2$$
\begin{equation}
e^{i(X+Y)}\frac{dXdY d \mu d\nu}{2\pi}=1.
\end{equation}

\section{Momentum representation}
\pst
Probability density to find particle in interval $(x,x+dx)$ is $|\psi(x)|^2$ and in interval $(p,p+dp)$ is $|\psi(p)|^2$,where $\psi(x)$and $\psi(p)$ are
wave functions in coordinate and momentum representations.They can be calculated through tomogram.
For delta potential (1) wave function in momentum representation is
\begin{equation}
\Psi(p)=\sqrt{\frac{2}{\pi}}\frac{\chi^{3/2}}{\chi^2+p^2}.
\end{equation}
We can find mean value of $p^2$ from (32)

\begin{equation}
<p^2>=\int p^2|\Psi(p)|^2dp =\chi^2.
\end{equation}

From formula (12) we have
$$M(X,\mu=0,\nu=1)=\int W(q,p)\delta(X-p)\frac{dqdp}{2\pi} =\int W(q,X)\frac{dq}{2\pi}=\left | \Psi (X) \right |^2$$
So in probability representation $<p^2>$ can be calculated from formula
\begin{equation}
<p^2>=\int_{-\infty }^{\infty}X^2 M(X,\mu=0,\nu=1)dX.
\end{equation}

It is easy to show from (4) that
$$M(X,\mu=0,\nu=1)=\frac{2\chi }{\pi}\frac{\chi^2}{(X^2+\chi^2)^2},$$
where we used asymptotic expansion of error function  $e^{z^2}erfc(z)=\frac{1}{z\sqrt{\pi}}+o(\frac{1}{z^3}) for z \to \infty.$

\begin{figure}[ht]
\bc \includegraphics[width=8.6cm]{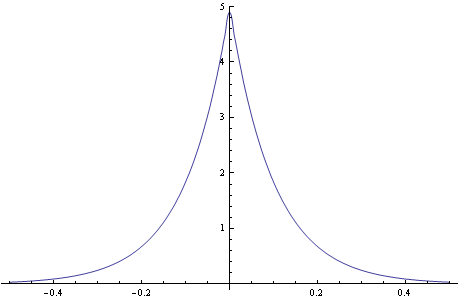}
 \includegraphics[width=8.6cm]{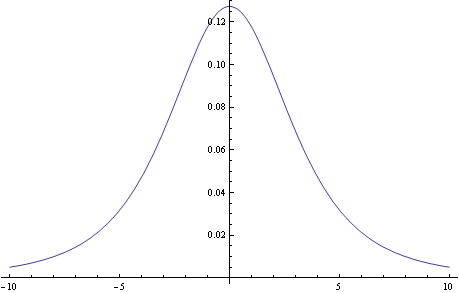} \ec
\vspace{-4mm}
\caption{
Optical tomograms $w(X,\theta \rightarrow 0)$ on fig ~(a) and $w(X,\theta \rightarrow \frac{\pi}{2})$ ~(b) for $\chi=5$. }
\end{figure}

We can calculate $<p^2>$ from (34)
\begin{equation}
\int_{-\infty }^{\infty} X^2 M(X,\mu=0,\nu=1)dX=\chi^2,
\end{equation}
which matches with (33).
From (4) it is obvious that for $\nu \to 0$ and $\mu = 1$
we have
$$M(X,\mu=1,\nu=0)=\chi e^{-2\chi|X|},$$
The figures 4 (a)and 4 (b) show  optical tomograms $w(X,\theta \rightarrow 0)$ and $w(X,\theta \rightarrow \frac{\pi}{2})$.
\begin{equation}
<x^2>=\int_{-\infty }^{\infty} X^2 M(X,\mu=1,\nu=0)dX=\frac{1}{2\chi^2}.
\end{equation}
From (35) and (36)
\begin{equation}
<\Delta x^2><\Delta p^2>=\frac{1}{2}.
\end{equation}

\section{Conclusion}
\pst
In this article  we found symplectic and optical tomograms for a particle moving in delta-potential.They are expressed through complex error function.
It has been shown that probability to remain in bound state when parameter of potential quickly changes can be calculated through the Wigner function or tomogram.
In addition we obtained the Wigner function for a particle moving in delta-potentials.It is worthy to note that the problem of the delta-potential barrier transparency was studied in \cite{22}.
Also problem of deflection and penetration of quantum particles in the case where the properties of delta-potential were taken into account was studied in \cite{23,24}  using error functions.

\section{Appendix 1}
\pst
In Appendix 1 we want to verify formula (28).$W_{1}(q,p)$and $W_{2}(q,p)$ are the Wigner functions,which correspond to the wave functions $\psi_{1}(x)$ and $\psi_{2}(x)$.
$$W_{1}(q,p)=\int du_{1}  e^{-ipu_{1} }\psi_{1} ^{*}(q+\frac{u_{1} }{2})\psi_{1} (q-\frac{u_{1} }{2}),$$
$$W_{2}(q,p)=\int du_{2}  e^{-ipu_{2} }\psi_{2} ^{*}(q+\frac{u_{2} }{2})\psi_{2} (q-\frac{u_{2} }{2})$$
Substitute $W_{1}(q,p)$ and $W_{2}(q,p)$ into (28)
$$P(0 \to 1)= \frac{1}{2\pi}\int  dq dp du_{1} du_{2} e^{-ip(u_{1}+u_{2}) }\psi_{1} ^{*}(q+\frac{u_{1} }{2})\psi_{1} (q-\frac{u_{1} }{2})  \psi_{2} ^{*}(q+\frac{u_{2} }{2})\psi_{2} (q-\frac{u_{2} }{2})$$
Using $\frac{1}{2\pi}\int_{-\infty }^{\infty } e^{ikx} dx=\delta(k)$
$$P(0 \to 1)=\int  dq  du \psi_{1} ^{*}(q-\frac{u }{2})\psi_{1} (q+\frac{u }{2})  \psi_{2} ^{*}(q+\frac{u }{2})\psi_{2} (q-\frac{u }{2})=$$
$$=\int  dx dy \psi_{1} ^{*}(x)\psi_{1} (y)  \psi_{2} ^{*}(y)\psi_{2} (x)=
\left |\int \psi_{1} (x) \psi_{2} ^{*}(x)dx\right |^2.$$

\section{Appendix 2}
\pst
In Appendix 2 we want to verify formula (29).
$M_{1}(X,\mu ,\nu )$ and $M_{2}(X,\mu ,\nu )$ are tomograms,which correspond to the Wigner functions $W_{1}(q,p)$and $W_{2}(q,p)$.
$$M_{1}(X,\mu,\nu)=\int W_{1}(q,p)\delta (X-\mu q - \nu p)\frac{dqdp}{2\pi },$$
$$M_{2}(X,\mu,\nu)=\int W_{2}(q,p)\delta (X-\mu q - \nu p)\frac{dqdp}{2\pi }$$
Substitute $M_{1}(X,\mu ,\nu )$ and $M_{2}(X,\mu ,\nu )$ into (29)
$$P(0 \to 1)=\int M_{1}(X,\mu ,\nu )M_{2}(Y,-\mu ,-\nu )e^{i(X+Y)}\frac{dXdY d \mu d\nu}{2\pi}=$$
$$\frac{1}{(2\pi)^3}\int  W_{1}(q_{1},p_{1})W_{2}(q_{2},p_{2})\delta (X-\mu q_{1} - \nu p_{1}) \delta (Y+\mu q_{2} + \nu p_{2})e^{i(X+Y)}d\mu d\nu dX dY dp_{1}dp_{2}dq_{1}dq_{2}$$
$$=\frac{1}{(2\pi)^3}\int  W_{1}(q_{1},p_{1})W_{2}(q_{2},p_{2})e^{i(\mu(q_{1}-q_{2})+\nu(p_{1}-p_{2}))}d\mu d\nu dp_{1}dp_{2}dq_{1}dq_{2}=
\frac{1}{2\pi}\int W_{1}(q,p)W_{2}(q,p)dqdp.$$

\section{Acknowledgments}
\pst This study was partially supported by the Russian Foundation for Basic Research by the Project RFBR 11-02-00456.

\end{document}